\documentclass[aps,prl,twocolumn,groupedaddress,showpacs]{revtex4}

\usepackage{graphicx}
\begin{document}

\title{Quantum phase interference and spin parity in Mn$_{12}$ single-molecule magnets}

\author{W. Wernsdorfer$^1$, N. E. Chakov$^2$, and G. Christou$^2$}


\affiliation{
$^1$Lab. L. N\'eel, associ\'e \`a l'UJF, CNRS, BP 166,
38042 Grenoble Cedex 9, France\\
$^2$Dept. of Chemistry, Univ. of Florida,
Gainesville, Florida 32611-7200, USA
}

\date{\today}

\begin{abstract}
Magnetization measurements of Mn$_{12}$ molecular nanomagnets 
with spin ground states of $S = 10$ and $S = 19/2$ show
resonance tunneling at avoided energy level crossings. 
The observed oscillations of the tunnel probability as a 
function of the magnetic field applied along the 
hard anisotropy axis are due to topological 
quantum phase interference of two tunnel paths of 
opposite windings. 
Spin-parity dependent tunneling is established by comparing
the quantum phase interference of integer and half-integer spin 
systems. 
\end{abstract}


\maketitle
Studying the limits between classical and 
quantum physics has become a very attractive 
field of research. Single-molecule magnets (SMMs) are 
among the most promising candidates to observe 
these phenomena since they have a well defined 
structure with well characterized spin ground state 
and magnetic anisotropy.
The first molecule shown to be a SMM 
was Mn$_{12}$acetate~\cite{Sessoli93b,Sessoli93}. 
It exhibits slow magnetization relaxation of its $S$ = 10 ground state 
which is split by axial zero-field splitting. It was the first system 
to show thermally assisted tunneling of magnetization
~\cite{Friedman96,Thomas96} and 
Fe$_8$ and Mn$_4$ SMMs were the first to exhibit ground state 
tunneling~\cite{Sangregorio97,Aubin98}. 
Tunneling was also found 
in other SMMs (see, for instance,~\cite{Caneschi99,Price99,Yoo_Jae00}). 
Quantum phase interference~\cite{Garg93} is among the most 
interesting quantum phenomena that can be studied
at the mesoscopic level in SMMs.
This effect was recently observed in Fe$_8$ and [Mn$_{12}$]$^{2-}$ 
SMMs~\cite{WW_Science99,WW_JAP02}.
It has led to new theoretical studies on quantum phase interference in spin systems 
~\cite{Garg99a,Garg99b,Garg99c,Garg00b,Garg00d,Barnes99,Villain00,Liang00,Yoo00,Yoo00b,Leuenberger00,Leuenberger01b,Lu00a,Lu00b,Lu00c,Zhang99,Jin00,Chudnovsky00a}.

The spin-parity effect is another
fundamental prediction which has rarely been observed at the 
mesoscopic level~\cite{WW_PRB02}. It predicts that quantum tunneling is 
suppressed at zero applied field if the total spin of the magnetic 
system is half-integer but is allowed in integer spin systems. 
Enz, Schilling, Van Hemmen and S$\ddot{\rm u}$to~\cite{Enz86,VanHemmen86} 
were the first to suggest the 
absence of tunneling as a consequence of Kramers degeneracy~\cite{note1}.
It was then shown that tunneling can even be absent without 
Kramers degeneracy~\cite{Loss92,Delft92,Garg93}; 
quantum phase interference can lead to 
destructive interference and thus 
suppression of tunneling~\cite{Garg93}. This 
effect was recently seen in Fe$_8$ 
and Mn$_{12}$ SMMs~\cite{WW_Science99,WW_JAP02}.

\begin{figure}
\begin{center}
\includegraphics[width=.3\textwidth]{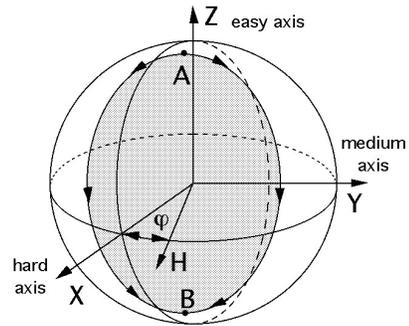}
\caption{Unit sphere showing degenerate minima 
{\bf A} and {\bf B} which are joined by two tunnel 
paths (heavy lines). The hard, medium, 
and easy axes are taken in $x$-, $y$-, and 
$z$-direction, respectively. 
The constant transverse field $H_{trans}$ for tunnel splitting
measurements is applied in 
the $xy$-plane at an azimuth angle 
$\varphi$. At zero applied field $\vec{H} = 0$, the giant spin 
reversal results from the interference of two 
quantum spin paths of opposite direction 
in the easy anisotropy $yz$-plane.
For transverse fields in the direction of the hard axis,
the two quantum spin paths are in a plane which is parallel to the 
$yz$-plane, as indicated in the figure.
By using Stokes'theorem it has been shown \cite{Garg93} that 
the path integrals can be converted into an area integral, 
leading to destructive interference---that is a 
quench of the tunneling rate---occurring whenever 
the shaded area is $k \pi / S$, where $k$ is an odd integer.
The interference effects disappear quickly when the transverse field 
has a component in the $y$-direction because the tunneling is then 
dominated by only one quantum spin path.}
\label{sphere}
\end{center}
\end{figure}

There are several reasons why 
quantum phase interference and spin-parity effects are
difficult to observe. The main reason reflects 
the influence of environmental degrees of freedom that can induce or 
suppress tunneling: hyperfine and dipolar couplings can 
induce tunneling via transverse field components;  
intermolecular superexchange coupling may enhance or suppress tunneling 
depending on its strength; phonons can induce transitions via excited 
states; and faster-relaxing species can complicate the 
interpretation~\cite{WW_EPL99}. 

We present here the first half-integer spin SMM that clearly shows 
quantum phase interference and spin-parity effects.
The syntheses, crystal structures and magnetic properties
of the studied complexes are reported elsewhere~\cite{Chakov05}.
The compounds are
[Mn$_{12}$O$_{12}$(O$_2$CC$_6$F$_5$)$_{16}$(H$_2$O)$_4$], 
(NMe$_4$)[Mn$_{12}$O$_{12}$(O$_2$CC$_6$F$_5$)$_{16}$(H$_2$O)$_4$], and 
(NMe$_4$)$_2$[Mn$_{12}$O$_{12}$(O$_2$CC$_6$F$_5$)$_{16}$(H$_2$O)$_4$] (called Mn$_{12}$, [Mn$_{12}$]$^{-}$, and 
[Mn$_{12}$]$^{2-}$, respectively).
Reaction of Mn$_{12}$ with one and two equivalents of NMe$_4$I 
affords the one- and two-electron reduced analogs, 
[Mn$_{12}$]$^{-}$ and [Mn$_{12}$]$^{2-}$, respectively. The three complexes 
crystallize in the triclinic P1bar, monoclinic P2/c 
and monoclinic C2/c space groups, respectively. 
The molecular structures are all very similar, each consisting 
of a central [Mn$^{\rm IV}$O$_4$] cubane core that is surrounded by a 
non-planar ring of eight Mn$^{\rm III}$ ions. 
Bond valence sum calculations establish that the added electrons 
in [Mn$_{12}$]$^{-}$ and [Mn$_{12}$]$^{2-}$ are localized on former Mn$^{\rm III}$ ions 
giving trapped-valence Mn$_4^{\rm IV}$Mn$_7^{\rm III}$Mn$^{\rm II}$
and Mn$_4^{\rm IV}$Mn$_6^{\rm III}$Mn$_2^{\rm II}$ anions, respectively. 

Magnetization studies yield 
$S = 10$, $D = 0.58$ K, $g = 1.87$ for Mn$_{12}$, 
$S = 19/2$, $D = 0.49$ K, $g = 2.04$, for [Mn$_{12}$]$^{-}$, and 
$S = 10$, $D = 0.42$ K, $g = 2.05$, for [Mn$_{12}$]$^{2-}$, 
where $D$ is the axial zero-field splitting parameter~\cite{Chakov05}. 
AC susceptibility and relaxation measurements
give Arrhenius plots from which were obtained the 
effective barriers to magnetization reversal: 
59 K for Mn$_{12}$, 49 K for [Mn$_{12}$]$^{-}$, and 25 K for [Mn$_{12}$]$^{2-}$.

\begin{figure}
\includegraphics[width=.45\textwidth]{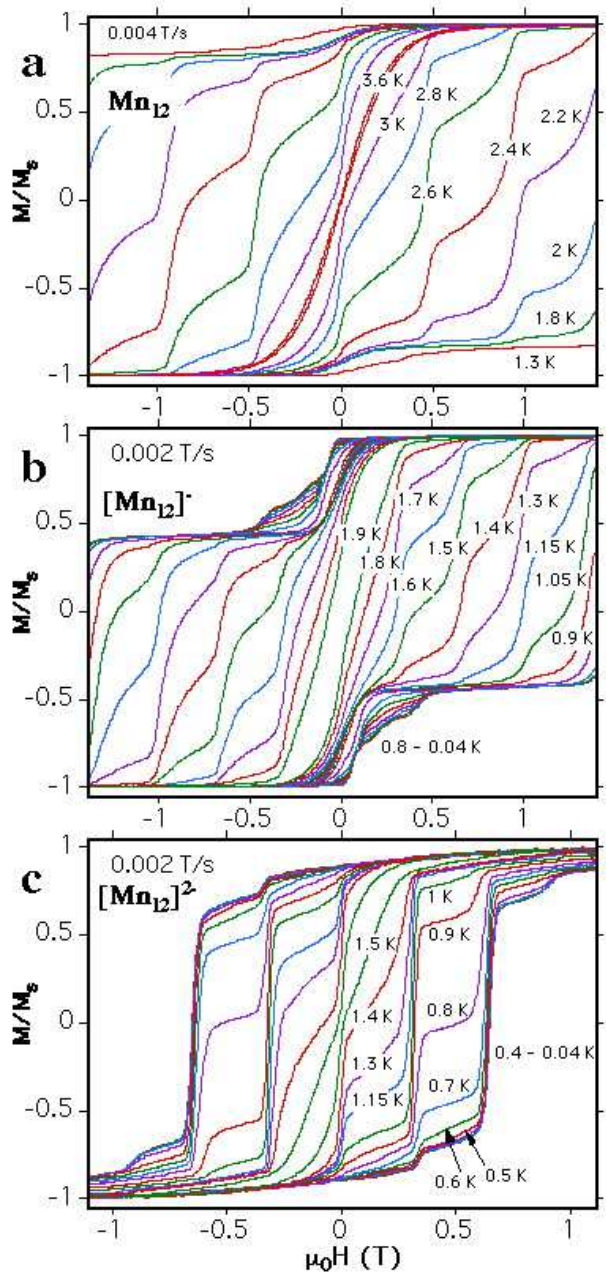}
\caption{Hysteresis loops of single crystals 
of (a) Mn$_{12}$, (b) [Mn$_{12}$]$^{-}$, and (c) [Mn$_{12}$]$^{2-}$ molecular clusters
at different temperatures and a constant field
sweep rate indicated in the figure. 
Note the large zero field step of [Mn$_{12}$]$^{-}$ which is
due to fast-relaxing species~\cite{note2}.}
\label{hyst}
\end{figure}

The simplest model describing the spin system of the three Mn$_{12}$ SMMs
has the following Hamiltonian
\begin{equation}
	H = -D S_z^2 + E \left(S_x^2 - S_y^2\right) 
	- g \mu_{\rm B} \mu_0 \vec{S}\cdot\vec{H} 
\label{eq_H_biax}
\end{equation}
$S_x$, $S_y$, and $S_z$ are the three 
components of the spin operator, 
$D$  and $E$ are the anisotropy constants, 
and the last term describes the Zeeman 
energy associated with an applied field $\vec{H}$. 
This Hamiltonian defines hard, medium, 
and easy axes of magnetization in $x$, $y$, and $z$ directions, 
respectively (Fig. 1). 
It has an energy level spectrum with $(2S+1)$ values which, 
to a first approximation, can be labeled by the 
quantum numbers $m = -S, -(S-1), ..., S$
taking the $z$-axis as the quantization axis. The energy spectrum  
can be obtained by using 
standard diagonalisation techniques of the $[(2S+1) \times (2S+1)]$ 
matrix. At $\vec{H} = 0$, the levels $m = \pm S$ 
have the lowest energy. 
When a field $H_z$ is applied, the levels with 
$m < 0$ increase in energy, while those 
with $m > 0$ decrease. Therefore, 
energy levels of positive and negative 
quantum numbers cross at certain values of $H_z$, 
given by $\mu_0 H_z \approx n D/g \mu_{\rm B}$, 
with $n = 0, 1, 2, 3, ...$. 

When the spin Hamiltonian contains transverse terms 
(for instance $E(S_x^2 - S_y^2)$), the level crossings  can be 
avoided level crossings. 
The spin $S$ is in resonance between two 
states when the local longitudinal field is close 
to an avoided level crossing. 
The energy gap, the so-called 
tunnel spitting $\Delta$, can be tuned by 
a transverse field (Fig. 1) 
via the $S_xH_x$ and $S_yH_y$ Zeeman terms.
In the case of the transverse term $E(S_x^2 - S_y^2)$,
it was shown that $\Delta$ oscillates with
a period given by \cite{Garg93}
\begin{equation}
\mu_0\Delta H = \frac {2 k_{\rm B}}{g \mu_{\rm B}} \sqrt{2 E (E + D)}
\label{eq_Garg}
\end{equation}
The oscillations are explained by constructive 
or destructive interference of quantum 
spin phases (Berry phases) of two tunnel paths \cite{Garg93} (Fig. 1).

All measurements were performed using an array of 
micro-SQUIDs~\cite{WW_ACP01}. The high sensitivity of this 
magnetometer allows the study of single crystals of SMMs
with sizes of the order of 10 to 500 $\mu$m.
The field can be applied in any direction by separately 
driving three orthogonal coils. The field was aligned using
the transverse field method~\cite{WW_PRB04}. 

Fig. 2 shows typical hysteresis loop 
measurements on a single crystal of the three Mn$_{12}$ samples. 
The effect of avoided level crossings can 
be seen in hysteresis loop measurements. 
When the applied field is near an 
avoided level crossing, the magnetization relaxes faster, 
yielding steps separated by plateaus. 
As the temperature is lowered, there is a decrease 
in the transition rate as a result of 
reduced thermally assisted tunneling.
Below about $T_{\rm c}$ = 0.65 K, 0.5 K, 0.35 K, respectively for
Mn$_{12}$, [Mn$_{12}$]$^{-}$, [Mn$_{12}$]$^{2-}$, the  hysteresis loops become 
temperature independent which suggests that the
ground state tunneling is dominating. 
The field between two resonances
allows an estimation of the anisotropy constants $D$,
and values of $D \approx $ 0.64 K, 0.44 K, 0.42 K were determined 
(supposing $g = 2$), 
respectively for Mn$_{12}$, [Mn$_{12}$]$^{-}$, [Mn$_{12}$]$^{2-}$, being in
good agreement with other magnetization studies~\cite{Chakov05}.

\begin{figure}
\begin{center}
\includegraphics[width=.45\textwidth]{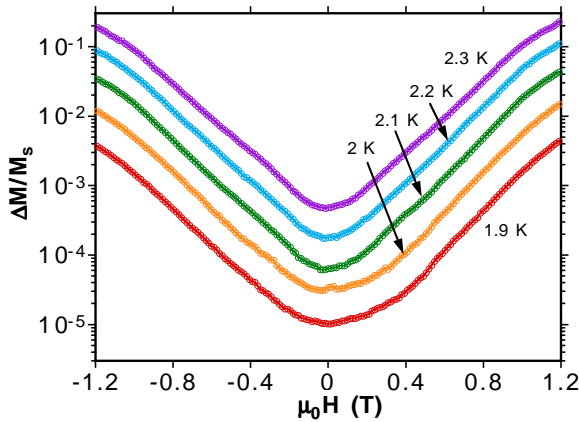}
\caption{Fraction of Mn$_{12}$ molecules which reversed their magnetization after
the  field was swept over the zero field resonance  at a rate of 0.28 
T/s  and at several temperatures.}
\label{P_Mn12_n}
\end{center}
\end{figure}

\begin{figure}
\begin{center}
\includegraphics[width=.45\textwidth]{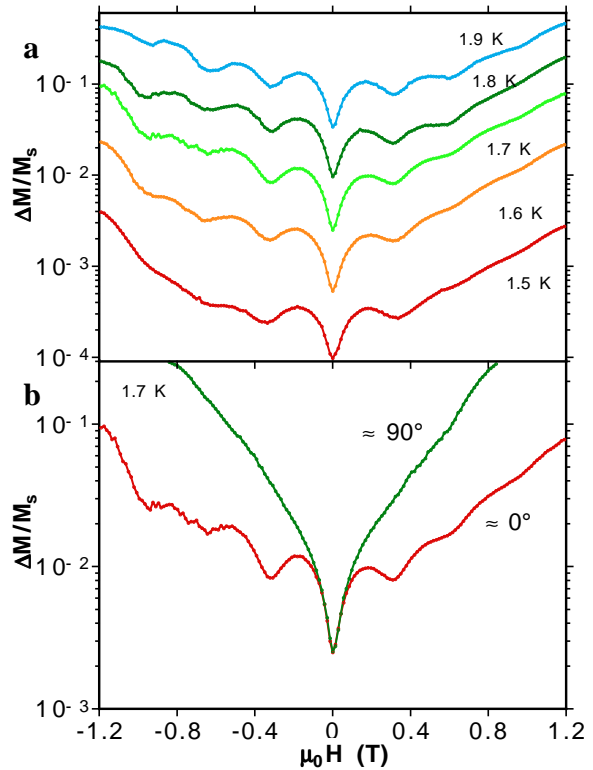}
\caption{Fraction of [Mn$_{12}$]$^{-}$ molecules which reversed their magnetization after
the  field was swept over the zero field resonance  at a rate of 0.28 
T/s (a) at several 
temperatures and (b) at 1.7 K and two  azimuth angles $\varphi$.
The contribution of the  fast-relaxing species
is substracted. The observed oscillations are 
direct evidence for quantum phase interference.
The minimum of the tunnel rate at zero transvers field is
due to Kramers spin parity.}
\label{P_Mn12_e}
\end{center}
\end{figure}

\begin{figure}
\begin{center}
\includegraphics[width=.45\textwidth]{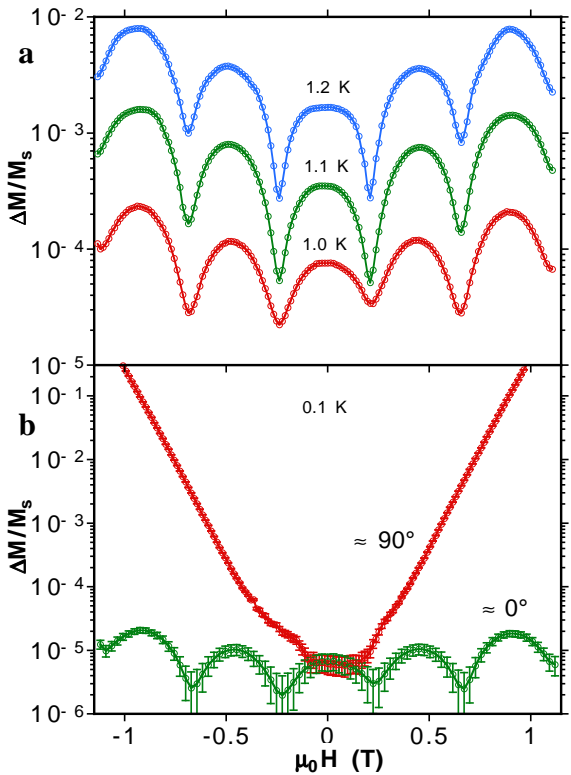}
\caption{Fraction of [Mn$_{12}$]$^{2-}$ molecules which reversed their magnetization after
the  field was swept over the zero field resonance at a rate of 0.28 
T/s (a) at several 
temperatures and (b) at 0.1 K and two  azimuth angles $\varphi$.}
\label{P_Mn12_2e}
\end{center}
\end{figure}

We have tried to use the
Landau--Zener method~\cite{Landau32,Zener32} to measure
the tunnel splitting as a function of transverse field
as previously reported for Fe$_8$~\cite{WW_Science99},. 
However, the tunnel probability in the pure quantum
regime (below $T_{\rm c}$) was too small for our measuring 
technique~\cite{note2} for Mn$_{12}$ and [Mn$_{12}$]$^{-}$.
We therefore studied the tunnel probability in
the thermally activated regime~\cite{WW_EPL00}.

In order to measure the tunnel probability,
the crystals of Mn$_{12}$ SMMs were first placed in a high 
negative field, yielding a saturated initial magnetization. 
Then, the applied field was swept at a constant rate 
of 0.28 T/s over the zero field resonance transitions and 
the fraction of molecules which 
reversed their spin was measured. In the case of
very small tunnel probabilities, the field was 
swept back and forth over the zero field resonance
until a larger fraction of molecules reversed their spin.
A scaling procedure yields the probability of one sweep.
This experiment was
then repeated but in the presence of a
constant transverse field. A typical result is
presented in Fig. 3 for Mn$_{12}$ showing a monotonic
increase of the tunnel probability. Measurements
at different azimuth angles $\varphi$ (Fig. 1) did
not show a significant difference. However, similar
measurements on [Mn$_{12}$]$^{-}$ (Fig. 4) and [Mn$_{12}$]$^{2-}$ (Fig. 5) showed oscillations 
of the tunnel probability as a 
function of the magnetic field applied along the 
hard anisotropy axis $\varphi = 0^{\circ}$ whereas no
oscillations are observed for $\varphi = 90^{\circ}$. 
These oscillations are due to topological 
quantum interference of two tunnel paths of 
opposite windings~\cite{Garg93}. The measurements  
of [Mn$_{12}$]$^{2-}$ are similar 
to the result on the Fe$_8$ molecular 
cluster~\cite{WW_Science99,WW_EPL00}; however,
those of [Mn$_{12}$]$^{-}$ show a minimum of the tunnel probability
at zero transverse field.
This is due to the spin-parity effect that predicts
the absence of tunneling as a consequence of Kramers degeneracy~\cite{note1}.
The period of oscillation
allows an estimation of the anisotropy constant $E$ (see Eq. 2) 
and values of $E \approx $ 0 , 0.047 K, and 0.086 K were obtained
for Mn$_{12}$, [Mn$_{12}$]$^{-}$, [Mn$_{12}$]$^{2-}$, respectively.

In conclusion, magnetization measurements of three molecular Mn$_{12}$ clusters 
with a spin ground state of $S = 10$ and $S = 19/2$ show
resonance tunneling at avoided energy level crossings. 
The observed oscillations of the tunnel probability as a 
function of a transverse field are due to topological 
quantum phase interference of two tunnel paths of 
opposite windings. 
Spin-parity dependent tunneling 
is established by comparing
the quantum phase interference of integer and half-integer spin 
systems. 

This work was supported by the EC-TMR Network 
ÒQuEMolNaÓ (MRTN-CT-2003-504880), CNRS and Rhone-Alpe funding.


\begin{thebibliography}{10}
\bibitem{note2} 
As observed for Mn$_{12}$ acetate~\cite{WW_EPL99},
the crystals of Mn$_{12}$, [Mn$_{12}$]$^{-}$, [Mn$_{12}$]$^{2-}$ contain a small fraction of
faster-relaxing species, which are probably molecules
having a defect. The signals of these species were very
large compared to the ground state relaxation rate
of the major species.
\end{thebibliography}

\end{document}